\documentclass[twocolumn,aps,pra,superscriptaddress,showpacs]{revtex4}
\usepackage{graphicx}
\usepackage{amsmath}
\usepackage{epsfig}
\usepackage{amssymb}

\begin{document}
\title{Discrimination of binary coherent states using a homodyne detector and a photon number resolving detector}

\author{Christoffer Wittmann}
\affiliation{Max Planck Institute for the Science of Light, G\"{u}nther-Scharowsky-Str. 1, Bau 24, 91058, Erlangen, Germany}
\affiliation{Institut f\"{u}r Optik, Information und Photonik, University of Erlangen-Nuremberg, Staudtstra\ss e 7/B2, 91058, Erlangen, Germany}

\author{Ulrik L. Andersen}
\affiliation{Department of Physics, Technical University of Denmark, Building 309, 2800 Kgs. Lyngby, Denmark}
\affiliation{Max Planck Institute for the Science of Light, G\"{u}nther-Scharowsky-Str. 1, Bau 24, 91058, Erlangen, Germany}

\author{Masahiro Takeoka}
\affiliation{National Institute of Information and Communications Technology (NICT),
4-2-1 Nukui-kitamachi, Koganei, Tokyo 184-8795, Japan}


\author{Gerd Leuchs}
\affiliation{Max Planck Institute for the Science of Light, G\"{u}nther-Scharowsky-Str. 1, Bau 24, 91058, Erlangen, Germany}
\affiliation{Institut f\"{u}r Optik, Information und Photonik, University of Erlangen-Nuremberg, Staudtstra\ss e 7/B2, 91058, Erlangen, Germany}

\date{\today}

\begin{abstract}
We investigate quantum measurement strategies capable of discriminating two coherent states probabilistically with significantly smaller error probabilities than can be obtained using non-probabilistic state discrimination. We apply a postselection strategy to the measurement data of a homodyne detector as well as a photon number resolving detector in order to lower the error probability. We compare the two different receivers with an optimal intermediate measurement scheme where the error rate is minimized for a fixed rate of inconclusive results. The photon number resolving (PNR) receiver is experimentally demonstrated and compared to an experimental realization of a homodyne receiver with postselection. In the comparison it becomes clear, that the perfromance of the new PNR receiver surpasses the performance of the homodyne receiver, which we proof to be optimal within any Gaussian operations and conditional dynamics. 
\end{abstract}


\maketitle
\section{Introduction}
In classical communication systems, information is usually encoded into orthogonal quantum or semi-classical states of light. An important example is binary phase shift keying with coherent states where the logical information, "0" and "1", is encoded as two coherent states with large amplitudes and a relative phase of $\pi$. Since the two coherent states each possess a large amplitude (with opposite signs), they are nearly orthogonal and thus they can be easily discriminated using an interferometric measurement strategy. On the other hand, if the amplitude becomes very small, which is the case for quantum communication as well as long distance amplification-free (e.g. free-space) classical communication, the two states are largely overlapping and thus non-orthognal. Due to this non-orthogonality, the states can no longer be perfectly discriminated. Although perfect discrimination is not possible, it is of high interest to find optimized strategies in order to minimize measurement errors, thus keeping the error rate as low as possible and increasing the mutual information between sender and receiver. Moreover, the search for such optimised strategies are of utmost importance for many applications in quantum communication with quantum key distribution (QKD) being the prime example~\cite{Bennett1984, Bennett1992, Grosshans2003}. Finally, we note that the problem of finding optimised measurement schemes associted with pre-defined alphabet is a fundamental problem in quantum mechnics~\cite{Helstrom1976, Ivanovic1987}.

There are basically two well-known discrimination strategies. In the first strategy, all measurement outcomes are used (that is, it is deterministic) and, therefore, the resulting conclusions will be infected by errors. The idea is to optimise the strategy such that the probability for making an error is minimized. This strategy is known as minimum error state discrimination and has been analyzed by Helstrom~\cite{Helstrom1976}. 
The second discrimination strategy is probabilistic, and yields only a valid outcome when the conclusion drawn from the measurement is known to be error-free. Therefore, in this task the goal is to minimise the probability of inconclusive results (which are discarded). This strategy is known as unambiguous state discrimination (USD) and was originally proposed by Ivanovic, Dieks and Peres~\cite{Ivanovic1987,Dieks1988, Peres1988, Jaeger1995}.
A combination of the two discrimination schemes where one allows for both erroneous and inconclusive results has also been treated theoretically. More precisely, the minimal probability of errors for a fixed probability of inconclusive results has been derived for pure and mixed states in refs.~\cite{Chefles1998} and~\cite{fiurek_optimal_2003}, respectively. 

For the discrimination of two coherent states with minimum error, several optimal and near-optimal receivers have been proposed~\cite{Kennedy1973, Dolinar1973, Sasaki1996,  Geremia2004, Takeoka2005,Takeoka2008}. 
Also, a device for implementing USD of coherent states was proposed by Huttner~\cite{Huttner1995} and later by Banaszek~\cite{Banaszek1999a}. Some of these schemes have been experimentally accessed, such as the Dolinar receiver~\cite{Cook2007}, the optimized displacement receiver~\cite{Wittmann2008c} and a programmable receiver implementing USD~\cite{bartkov_fiber-optics_2007}. However, the intermediate regime where errors as well as  inconclusive results may occur has only recently been investigated experimentally~\cite{Wittmann2009d}. 

In this paper we elaborate on the work of ref~\cite{Wittmann2009d}. We investigate two different receivers that belong to the intermediate regime. The first receiver is a standard homodyne detector, and the second receiver is a displacement controlled photon number resolving detector~\cite{Wittmann2009}. In both receivers the measurement outcomes are postselected to obtain a specific relation between errors and inconclusive results. The postselection based homodyne detector has been used in various protocols such as QKD~\cite{Silberhorn2002, Lorenz2004, Lance2005}, squeezed state and entangled state distillation~\cite{fiurek_experimentally_2007,Heersink2006,Dong2008,Franzen2006,Hage2008} and quantum state engineering\cite{Marek2009, Lance2006, Ourjoumtsev2007, Babichev2004}. Here we thoroughly characterize the detector in terms of the discrimination between two coherent states. In addition we conduct a thorough experimental analysis of the displacement based photon number resolving detector which was introduced in ref~\cite{Wittmann2009, Wittmann2009d}. We find that the displacement based PNR receiver outperforms the standard homodyne detection.

The paper is organized as follows. First we recapitulate the notion of intermediate measurement for coherent states in section~\ref{measurement}. In section~\ref{HD} and section~\ref{PNR}, we consider measurements in the intermediate regime with two strategies: a receiver using a homodyne detection and a receiver using an optimized displacement combined with a photon number resolving detector. In section~\ref{HD}, we also prove that the post-selection based homodyne scheme is 
the optimal strategy for realizing the intermediate measurement 
within all possible Gaussian operations and conditional dynamics. We demonstrate experimentally both receivers in section~\ref{exp}.  Finally, we summarize the results in section~\ref{comp}.

\section{Intermediate measurement}
\label{measurement}

Let us assume a sender picks one signal state out of two pure and phase shifted coherent states and sends it through a communication channel that preserves the quantum property of the state. On the other end of the channel, a receiver has to tell which state was chosen by the sender. Let us also assume that the \textit{a priori} probabilities for the preparation of the states are $p_1=p_2=\frac{1}{2}$ and that the received states are $|-\alpha\rangle$ or $|+\alpha\rangle$.

The receiver measures the signal state and based on the measurement outcome guesses the state. Due to the non-orthogonality of the alphabet, however, the result will not be correct in all such attempts. In fact, the minimal error probability is given by the inner product of the states in the alphabet, $\sigma=|\langle-\alpha|\alpha\rangle|$. The maximally accessible information of the receiver is directly related to the minimal error rate.

The receiver can alternatively choose a measurement strategy which allows for inconclusive results. In this strategy, he will only accept states that are likely to be correctly identified, while he does not attempt to guess the results for signals associated with the inconclusive measurement results. This strategy is probabilistic as the outcomes are post selected. It can be shown, that for higher probability of inconclusive results $p_\mathrm{inc}$ (or lower acceptance probability $1-p_\mathrm{inc}$) a lower error probability $p_\mathrm{E}$ can be achieved. 

This intermediate measurement strategy can be described by the three-component positive operator-valued measure (POVM) $\hat\Pi_i, i=1,2,?$ where $\hat\Pi_i > 0$ and $\hat\Pi_1+\hat\Pi_2+\hat\Pi_?=\hat I$. Consequently, an inconclusive result will occur with the probability 
\begin{equation}
p_\mathrm{inc}=p_1\langle-\alpha|\hat\Pi_?|-\alpha\rangle+p_2\langle \alpha|\hat\Pi_?|\alpha\rangle.
\label{inconclusive}
\end{equation}
where $\langle-\alpha|\hat\Pi_?|-\alpha\rangle$ ($\langle \alpha|\hat\Pi_?|\alpha\rangle$) represents the  probability of inconclusive results when $|-\alpha\rangle$ $(|\alpha\rangle)$ was prepared.
Furthermore, the average error probability is given by
\begin{equation}
p_\mathrm{E}=\frac{p_1\langle-\alpha|\hat\Pi_2|-\alpha\rangle+p_2\langle \alpha|\hat\Pi_1|\alpha\rangle}{1-p_\mathrm{inc}},
\label{errorrate}
\end{equation}
where $\langle-\alpha|\hat\Pi_2|-\alpha\rangle$ ($\langle\alpha|\hat\Pi_1|\alpha\rangle$) represents the error probability of mistakenly guessing $|\alpha\rangle$ $(|+\alpha\rangle)$. 

Finally the measurement strategy is optimized, such that the receiver's error probability is minimized for a given probability of inconclusive results. The error probability according to~\cite{Chefles1998} is then given by
\begin{equation}
p_\mathrm{E}\ge \frac{1}{2} \left(1-\frac{\left(1-2p_\mathrm{inc}(1-\sigma)-\sigma^2\right)^{1/2}}{1-p_\mathrm{inc}}\right),
\label{opterror}
\end{equation}
where the error rate is lower bounded by the inner product and the tolerated rate of inconclusive results. A receiver scheme achieving this optimal bound is yet unknown. (To our knowledge this is also true for two non orthogonal qubits.) In the following two sections, we investigate two near-optimal receivers; the post-selection based homodyne receiver and displacement controlled photon number resolving detector.

\section{Homodyne receiver}
\label{HD}

\begin{figure}
\begin{tabular}{l}
\centerline{\includegraphics[width=8.4cm]{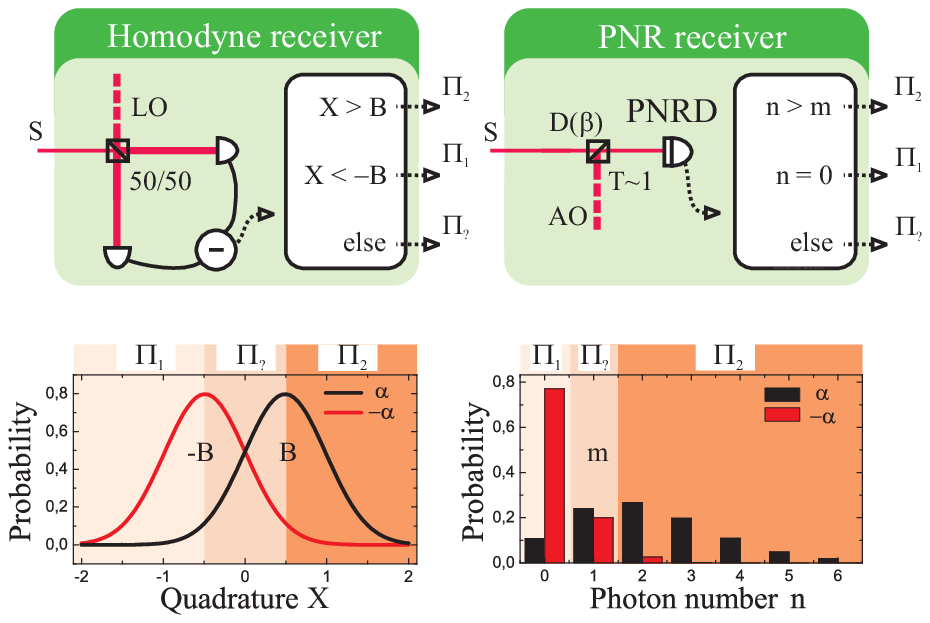}} \\ [-5.7cm]
(a)\hspace{3.6cm}(c) \\[2.4cm]
(b)\hspace{3.6cm}(d) \\[2.4cm]
(e)\\[-0.2cm]
\centerline{\includegraphics[width=8cm]{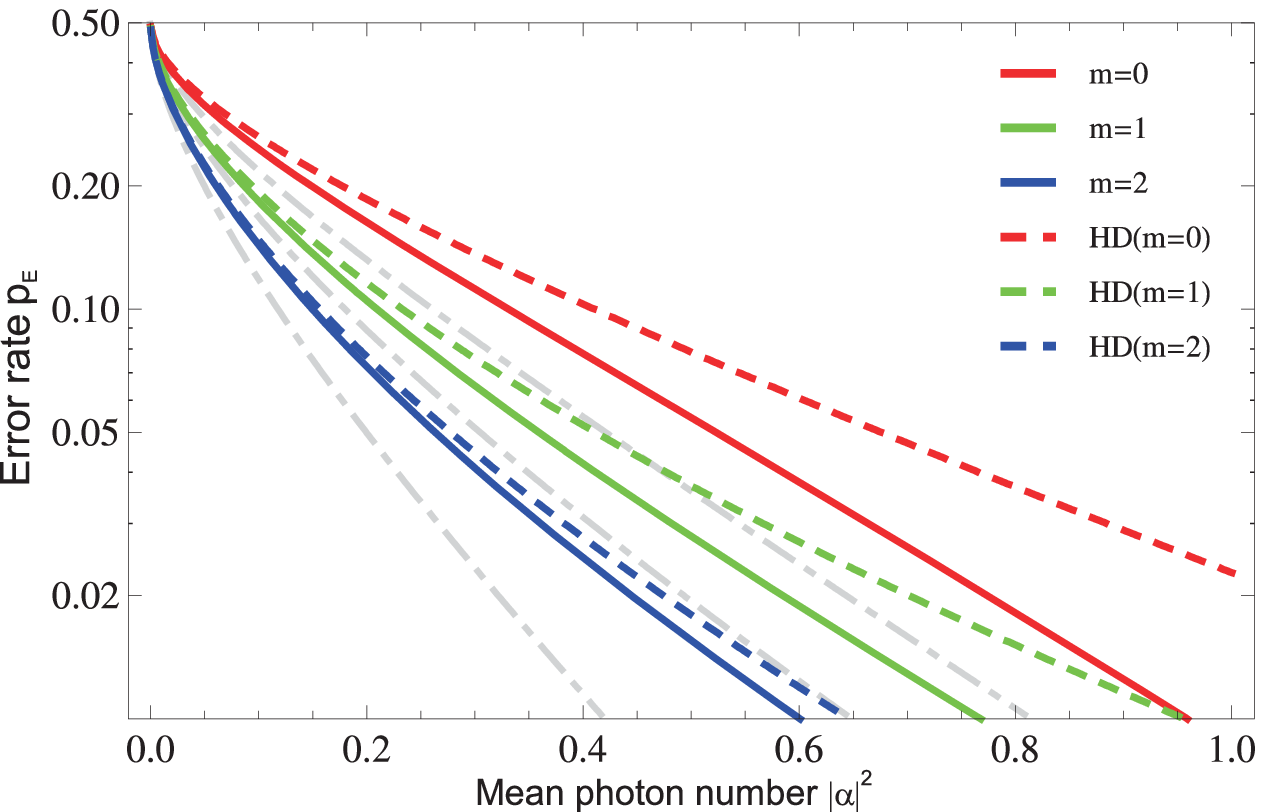}} \\
\end{tabular}
\caption{\label{schemes} (a) Schematics of the homodyne receiver, showing that the signal (S) is interfered with a local oscillator (LO) on a 50/50 beam splitter. The photocurrents of two photo diodes are subtracted and results in a quadrature measurement along the encoding quadrature. (b) Marginal distribution of the two signal states. In the example, we assume a signal with $|\alpha|^2=0.24$. According to the measurement result the correct answer ($-, ?, +$) is guessed. (c) Schematics of the photon number resolving (PNR) receiver. The signal (S) is interfered with an auxiliary oscillator (AO) on a highly transmissive beam splitter. Finally, the signal is measured by a photon number resolving detector (PNRD) (d) Photon number distribution of two signal states.  According to the measurement outcome of the PNRD the correct answer is guessed. In the example, we assume a signal with $|\alpha|^2=0.24$ and a displacement of $\beta=1$. Results of $n=1$ are considered inconclusive results. (e) Comparison of PNR receivers with $m=1$ to $3$ (solid lines) with homodyne receiver (dashed lines) at equal success rates. PNR receiver outperforms the homodyne receiver for all signal amplitudes. The dotdashed lines show the error rate of the optimal measurement.}
 
\end{figure}

A very simple receiver type, which is tunable in the probability of inconclusive results, is based on homodyne detection followed by postselection~\cite{Silberhorn2002, Lorenz2004, Lance2005}. The schematic of this receiver is shown in Fig.~\ref{schemes}(a). We now revisit homodyne detection with postselection in the context of state discrimination in the aforementioned intermediate regime.

In the homodyne measurement, the local oscillator is set along the excitation of the coherent states resulting in a distribution of quadrature values as shown in  Fig.~\ref{schemes}(b). The conclusion of the receiver is deduced from the particular result, where positive measurement outcomes greater than a postselection threshold $B$ identify $|\alpha\rangle$ whereas negative outcomes less that the postselection threshold $-B$ identify $|-\alpha\rangle$. All measurement outcomes between the postselection thresholds are considered as inconclusive results. The corresponding POVMs are $\hat\Pi_1=\int_{-\infty}^{-B}|x\rangle\langle x|dx$, $\hat\Pi_2=\int_B^\infty|x\rangle\langle x|dx$ and $\hat\Pi_?=\hat I-\hat\Pi_1-\hat\Pi_2$, and result in the error probability
\begin{equation}
p_{\mathrm{E,HD}}=\frac{1-\mathrm{erf}(\sqrt{2}(B+|\alpha|))}{2(1-p_\mathrm{inc,HD})}.
\label{pH}
\end{equation}
The probability of inconclusive results is found to be
\begin{equation}
p_{\mathrm{inc,HD}}=\frac{1}{2}\left(\mathrm{erf}(\sqrt{2}(B+|\alpha|))+\mathrm{erf}(\sqrt{2}(B-|\alpha|))\right).
\label{pHinc}
\end{equation}

In the following, we prove that the post-selected homodyne scheme is 
the optimal strategy for realizing the intermediate measurement 
within all possible Gaussian operations and conditional dynamics 
(classical feedback or feedforward). 
For simplicity we assume $\alpha$ is real and positive. 
Note that if the input alphabet as well as all operations 
are Gaussian, conditional dynamics is useless 
\cite{Eisert,Fiurasek,Giedke2002}. 
In our case, however, the input alphabet consists of an ensemble 
of two coherent states; $p_1 |{-}\alpha\rangle\langle{-}\alpha| 
+ p_2 |\alpha\rangle\langle\alpha|$. 
This is clearly non-Gaussian and thus we cannot discard 
conditional dynamics as a tool to improve the discrimination task. 
We first briefly introduce the characteristic functional formalism 
of POVMs and then discuss the POVMs via 
Gaussian operations with and without conditional dynamics.


Here we use a characteristic function formalism similar to the approach used to prove the optimality of the homodyne measurement 
for discriminating binary coherent states with minimum error 
under Gaussian operations \cite{Takeoka2008}. 
In quantum optics, the characteristic function $\chi(\omega)$ is often 
a useful tool to represent a continuous variable quantum state 
\cite{Walls1994}. 
In an $N$-mode infinite dimensional systems, the characteristic 
function of a quantum state with the density matrix $\hat{\rho}$ 
is defined as 
\begin{equation}
\label{eq:ch_func_state}
\chi_\rho(\omega) \equiv {\rm Tr}\left[ \hat{\rho} 
 \exp \left[ i \omega^T \hat{R} \right] \right], 
\end{equation}
where $\omega \in \mathbb{R}^{2N}$ and 
$\hat{R}= [\hat{x}_1, \cdots, \hat{x}_N, \hat{p}_1, \cdots, \hat{p}_N]$. 
Here $\hat{x}_i$ and $\hat{p}_i$ are the quadrature operators of 
mode $i$. 
In particular, a Gaussian state is defined as the state whose 
characteristic function is represented by a Gaussian function \cite{Ferraro2005};
\begin{equation}
\label{eq:ch_func_Gaussian_state}
\chi_\rho(\omega) = \exp \left[ -\frac{1}{4} \omega^T \Gamma \omega 
+ i D^T \omega \right], 
\end{equation}
where $\Gamma$ is a $2N \times 2N$ covariance matrix 
and 
$D$ is a $2N$ vector corresponding to the displacement.

A similar formalism is applicable for the representation of POVMs. 
A single-mode POVM consisting of any Gaussian operation, 
Gaussian auxiliary states, and homodyne measurements 
can be described by a set of operators 
$\{ \hat{\Pi}(d_\mathcal{M}) \}_{d_\mathcal{M}}$ whose 
characteristic function is \cite{Takeoka2008}
\begin{eqnarray}
\label{eq:ch_func_POVM}
\chi_d (\omega) & = & {\rm Tr}\left[ \hat{\Pi}(D) 
\exp \left[ i \omega^T \hat{R} \right] \right] 
\nonumber\\ 
& = & 
\frac{1}{\pi} \exp \left[ -\frac{1}{4} 
\omega^T \Gamma_\mathcal{M} \omega 
+ i d_\mathcal{M}^T \omega \right] ,
\end{eqnarray}
where $\Gamma_\mathcal{M}$ is a $2\times2$ covariance matrix and 
$d_\mathcal{M}=[u,v]^T$ representing the measurement outcome. 
A typical example of the measurement in this class is a heterodyne measurement 
described by $\{ \hat{\Pi}(\beta) = |\beta\rangle\langle\beta|/\pi 
\}_{\beta \in \mathbb{C}}$ whose covariance matrix is calculated 
to be an identity matrix and 
$d_\mathcal{M}=[\sqrt{2}{\rm Re}\beta, \sqrt{2}{\rm Im}\beta]^T$. 
Another example may be a homodyne measurement: 
A homodyne measurement with the phase $\varphi=0$ is 
a projection measurement onto an X-quadrature and  
its covariance matrix is given by 
$\Gamma_\mathcal{M} = {\rm diag}[e^{-2r}, e^{2r}]$ with $r\to \infty$ 
and the element in the first row of $d_\mathcal{M}$ corresponds to 
the measurement outcome. 
Note that the POVM in Eq.~(\ref{eq:ch_func_POVM}) does not 
include conditional dynamics. 
In this formalism, the probability distribution 
of detecting a state $\hat{\rho}$ by a POVM $\hat{\Pi}(d_\mathcal{M})$ is 
calculated as 
\begin{eqnarray}
\label{eq:prob_distribution}
P(d_\mathcal{M}) & = & {\rm Tr} \left[ \hat{\rho} 
\hat{\Pi}(d_\mathcal{M}) \right] 
\nonumber\\ 
& = & 
\frac{1}{2\pi} \int d\omega \chi_\rho (\omega) \chi_d (-\omega). 
\end{eqnarray}
More general characteristics of Gaussian state transformations 
in the formalism of characteristic function are described, e.g. 
in \cite{Giedke2002,Ferraro2005}


Let us construct the intermediate measurement 
via a Gaussian measurement 
described in Eq.~(\ref{eq:ch_func_POVM}), 
i.e. without conditional dynamics, 
and classical post-processing for a set of binary coherent states 
$\{ |\alpha\rangle, |{-}\alpha\rangle \}$. 
When the measurement is ``noise-free'' (i.e. 
consisting of Gaussian unitary operations, pure Gaussian ancillary 
states, and ideal homodyne measurements) 
the covariance matrix is simply given by~\cite{Takeoka2008},
\begin{eqnarray}
\label{eq:cov_matrix_Gaussian}
\Gamma_\mathcal{M} = \left[ 
\begin{array}{cc}
\cosh 2r - \sinh2r \cos\varphi & \sinh2r \sin\varphi \\
\sinh2r \sin\varphi & \cosh 2r + \sinh 2r \cos\varphi 
\end{array}
\right] , \nonumber\\
\end{eqnarray}
where $r$ and $\varphi$ are real parameters. 
This noise-free restriction does not compromise generality 
since one can always construct a general Gaussian measurement 
by preparing a corresponding noise-free Gaussian measurement and 
discarding a part of its measurement outcomes. 
We also note that such a noise-free Gaussian measurement corresponds to 
$\{ \frac{1}{\pi}|\psi_\zeta (u,v)\rangle\langle\psi_\zeta (u,v)| \}_{(u,v)}$, 
where $|\psi_\zeta(u,v)\rangle = \hat{D}(u,v) \hat{S}(\zeta) |0\rangle$, and 
$\hat{D}(u,v) = \exp[i(v\hat{x}-u\hat{p})]$ and 
$\hat{S}(\zeta) = \exp[(\zeta^* \hat{a}^2 - \zeta \hat{a}^{\dagger\,2})]$
are a displacement and a squeezing operators, respectively, 
and $\zeta=re^{i\varphi}$ is a complex squeezing parameter.

The characteristic functions of the coherent states $|\pm\alpha\rangle$ 
are given by $\chi_\pm (\omega) = \exp [-\frac{1}{4} \omega^T I
 \omega + i d_\pm^T \omega]$, where $I$ is the identity matrix and 
$d_\pm = [\pm\sqrt{2}\alpha, 0]^T$. 
The probability distribution of detecting $|\pm\alpha\rangle$ 
with such a POVM is thus calculated to be 
\begin{eqnarray}
\label{eq:prob_dist}
P(u,v|\pm) & = & 
\frac{1}{2\pi} \int d\omega \chi_\pm(\omega) \chi_D (-\omega) 
\nonumber\\ & = & 
\frac{1}{\pi \sqrt{\det (\Gamma_\mathcal{M} + I)}} 
\exp\left[ -\left\{ (\pm\sqrt{2}\alpha - u)^2 a \right. \right. 
\nonumber\\ && \left. \left. 
- 2 (\pm\sqrt{2}\alpha - u) v c + v^2 b \right\} \right], 
\end{eqnarray}
where 
\begin{equation}
\label{eq:gamma_inv}
\frac{1}{\Gamma_\mathcal{M} + I} = 
\left[
\begin{array}{cc}
a & c \\
c & b 
\end{array}
\right], 
\end{equation}
and 
\begin{eqnarray}
\label{eq:abc}
a & = & \frac{1+\cosh 2r+\sinh 2r \cos\varphi}{
   2(\cosh 2r+1)}, \\
b & = & \frac{1+\cosh 2r-\sinh 2r \cos\varphi}{
   2(\cosh 2r+1)}, \\
c & = & \frac{-\sin 2r \sin\varphi}{
   2(\cosh 2r+1)}. 
\end{eqnarray}

We first show that the optimal measurement in this class 
is homodyne detection.
Let us denote the likelihood ratio of two signals as 
\begin{equation}
\label{eq:likelihood}
\Lambda_1 = \frac{p_1 P(u,v|{-}\alpha)}{p_2 P(u,v|\alpha)}
= \frac{p_1}{p_2} \exp \left[-4\sqrt{2} \alpha (au+cv) \right] ,
\end{equation}
and $\Lambda_2{=}\Lambda_1^{-1}$. 
According to the Bayesian strategy \cite{Helstrom1976}, 
an optimal signal decision for the fixed measurement is to guess 
$|{-}\alpha\rangle$ for $\Lambda_1 {\ge} \Lambda_B$, 
$|\alpha\rangle$ for $\Lambda_2 {\ge} \Lambda_B$, 
and the inconclusive result otherwise, 
where $\Lambda_B$ is the threshold. 
The probabilities of having successive, erroneous, and 
inconclusive results for each signal are then given by 
\begin{eqnarray}
\label{eq:p_succ_E_inc}
p_s^{(\pm)} & = & \frac{1}{2} {\rm erfc}\left[ \sqrt{2a}\alpha 
- \frac{\ln\Lambda_B \pm \ln(p_1/p_2)}{4\sqrt{2}\alpha} \right], \\
p_e^{(\pm)} & = & \frac{1}{2} {\rm erfc}\left[ \sqrt{2a}\alpha 
+ \frac{\ln\Lambda_B \mp \ln(p_1/p_2)}{4\sqrt{2}\alpha} \right], \\
p_i^{(\pm)} & = & p_s^{(\pm)} - p_e^{(\pm)} . 
\end{eqnarray}
The average error and inconclusive probabilities are given by 
$p_E {=} (p_1 p_e^{(-)} {+} p_2 p_e^{(+)})/(1{-}p_{\rm inc})$ 
and $p_{\rm inc} {=} p_1 p_i^{(-)} {+} p_2 p_i^{(+)}$, respectively. 
We find that, for a given $\Lambda_B$, these two probabilities are 
simultaneously minimized for $\varphi=0$ and $r\to\infty$, i.e. 
an ideal homodyne measurement with phase $\varphi {=} 0$. 
This implies that the optimal measurement with only Gaussian operations 
is the homodyne detector with a fixed phase of $\varphi {=} 0$.


Furthermore, in the follwing we prove that any conditional operation will not improve the discrimination task. To prove this, we consider two different Gaussian operators. 
The first operation is a partial measurement of the signal which in general outputs 
a measurement outcome (classical number) and a conditioned output state. 
For an input of 
$p_1 |{-}\alpha\rangle\langle{-}\alpha| + p_2 |\alpha\rangle\langle\alpha|$, 
the conditioned output is given as
\begin{equation}
\label{eq:cond_output}
\hat{\rho}_{\rm out} = p'_1(d_\mathcal{M}) \hat{\rho}_- 
+ p'_2(d_\mathcal{M}) \hat{\rho}_+ ,
\end{equation}
where $\hat{\rho}_\pm$ are general multi-mode states that preserve 
Gaussian statistics with the joint covariance 
matrix $\Gamma_{\rm out}$ and the displacement $\pm D$. 
Here $d_\mathcal{M}$ denotes the measurement outcome and thus 
only the posteriori probabilities in Eq.~(\ref{eq:cond_output}) 
depend on $d_\mathcal{M}$. 
Moreover, it was shown that $\hat{\rho}_{\rm out}$ can be always 
transformed into 
another mixture of coherent states \cite{Takeoka2008}, 
\begin{equation}
\label{eq:cond_output2}
\hat{\rho}_{\rm out} \to \hat{\rho}_{\alpha'} \otimes \hat{\rho}_{\rm aux}, 
\end{equation}
where $\hat{\rho}_{\rm aux}$ is some Gaussian state and 
\begin{equation}
\label{eq:cond_output3}
\hat{\rho}_{\alpha'} = 
p'_1(d_\mathcal{M}) |{-}\alpha'\rangle\langle{-}\alpha'| + 
p'_2(d_\mathcal{M}) |\alpha'\rangle\langle\alpha'| ,
\end{equation}
with real and positive $\alpha'$. 
Such an additional Gaussian operation can be deterministic and 
independent of the partial measurement outcome $d_{\mathcal{M}}$. 
Since only the posteriori probabilities depend on $d_{\mathcal{M}}$, 
the optimal second operation 
is independent of $d_\mathcal{M}$ and given by 
a fixed homodyne measurement ($\varphi{=}0$) as already shown. 
We therefore conclude that any conditional dynamics is not useful 
in the two-step measurement scenario. 
An extension of the above conclusion to the multi-step measurement scenario 
is straightforward, which proves the optimality of the homodyne detector 
within all possible Gaussian operations and conditional dynamics. 

\section{Photon number resolving receiver}
\label{PNR}

Quadrature measurements (measurements of the light's field amplitude) and photon counting measurements (measurements of the excitation of a light field) are fundamentally different. Therefore, it is of interest to investigate also a receiver based on the latter technique. In Ref.~\cite{Wittmann2009} we proposed to use a photon number resolving (PNR) receiver for the discrimination of two coherent states. It consists of two stages: a displacement operation $D(\beta)$ and a photon number resolving detector (PNRD), and it is sketched in Fig.~\ref{schemes}(c).

The post selection process of the PNR receiver is similar to the one of the homodyne detector: If the measurement outcome of the PNR detector is $n=0$, we guess $|-\alpha\rangle$, if it is $n>m$ (where $m$ is the threshold parameter), we guess $|\alpha\rangle$ and otherwise the measurment is inconclusive. This can be described by the projector $\hat\Pi_?=\sum_{n=1}^{m}{|n\rangle\langle n|}$ for $m>0$. Conclusive results are described by $\hat\Pi_1=|0\rangle\langle 0|$ and $\hat\Pi_2=\hat I -\hat\Pi_1-\hat\Pi_?$, where $\hat\Pi_1$ identifies $|-\alpha\rangle$ and $\hat\Pi_2$ identifies $|\alpha\rangle$. An example for the photon number distributions of two displaced coherent states is shown in Fig.~\ref{schemes}(d). The error rate is then given by
\begin{equation}
p_{\mathrm{E,PNR}}=\frac{\left(1-\frac{\Gamma \left(\text{m}+1,(\alpha -\beta )^2\right)}{\Gamma (\text{m}+1)}+ e^{-(\alpha +\beta )^2}\right)}{2(1-p_{\mathrm{D,inc}})},
\label{pPNR}
\end{equation}
where the Euler gamma funtion $\Gamma(z)$ and the incomplete gamma function $\Gamma(a,z)$ are defined as $\Gamma (z)=\int _0^{\infty }t^{z-1}e^{-t}dt$ and $\Gamma (a,z)=\int _z^{\infty }t^{a-1}e^{-t}dt$. The probability of inconclusive results is given by
\begin{eqnarray}
\label{pPNRinc}
p_{\mathrm{inc,PNR}}=\frac{ \Gamma \left(m{+}1,(\alpha {-}\beta )^2\right)+ \Gamma \left(m{+}1,(\alpha {+}\beta )^2\right)}{2\Gamma (m{+}1)} \\
- \frac{1}{2} e^{-(\alpha -\beta )^2}-\frac{1}{2} e^{-(\alpha +\beta )^2}.\nonumber
\end{eqnarray}

The displacement in the receiver can be chosen such that one of two input states is displaced to the vacuum state ($\beta=\alpha$) as suggested by Kennedy~\cite{Kennedy1973}. However, to minimize the error rate of the receiver the displacement must be optimized, i.e. $\mathrm{d}p_{\mathrm{E,PNR}}/\mathrm{d}\beta = 0$. A detailed discussion of this receiver can be found in Ref.~\cite{Wittmann2009}.

We compare the PNR receiver with homodyne receivers with different postselection thresholds. In this comparison, we choose the postselection parameter $B$ such that the rates of inconclusive results are equal for both strategies, i.e. $p_\mathrm{inc,HD}=p_\mathrm{inc,PNR}$.  The error probability for the receivers with $m=0$ to $2$ and the corresponding homodyne receivers are plotted against the mean photon number of the signal in Fig.~\ref{schemes}(e). We find that the performance of the PNR receiver (solid lines) surpasses the performance of the homodyne receiver (dashed lines) for all signal amplitudes. The optimal discrimination strategy is shown by the grey) curve.

\section{Experimental Results}
\label{exp}

\begin{figure}
\begin{tabular}{l}
\centerline{\includegraphics[width=7.8cm]{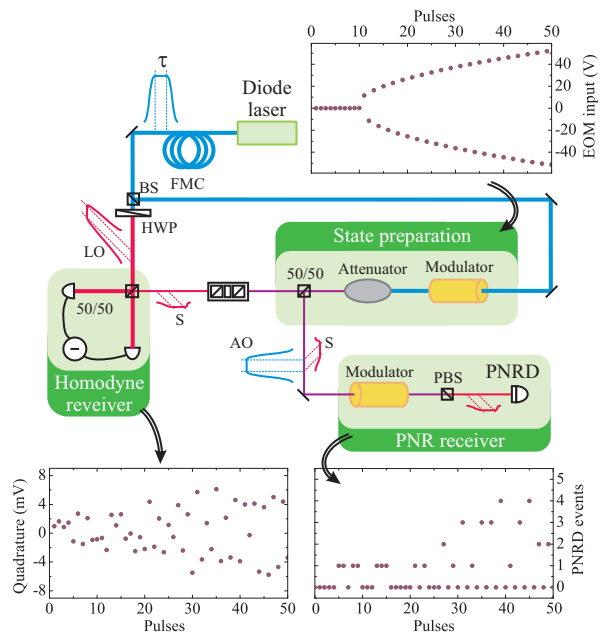}}\\[-0.1cm]
\end{tabular}
\caption{\label{SetupRaw} Simplified scheme of the experiment, where the abbreviated components are a fibre mode cleaner(FMC), beam splitters (BS, 50/50), a polarizing beam splitter (PBS), a half wave plate (HWP) and a photon number resolving detector (PNRD). The graphs show modulation and the corresponding recorded quadrature measurements and detection events of the PNRD.}
 
\end{figure}

\begin{figure*}
\begin{tabular}{ll}
\multicolumn{2}{c}{\includegraphics[width=0.7\textwidth,keepaspectratio]{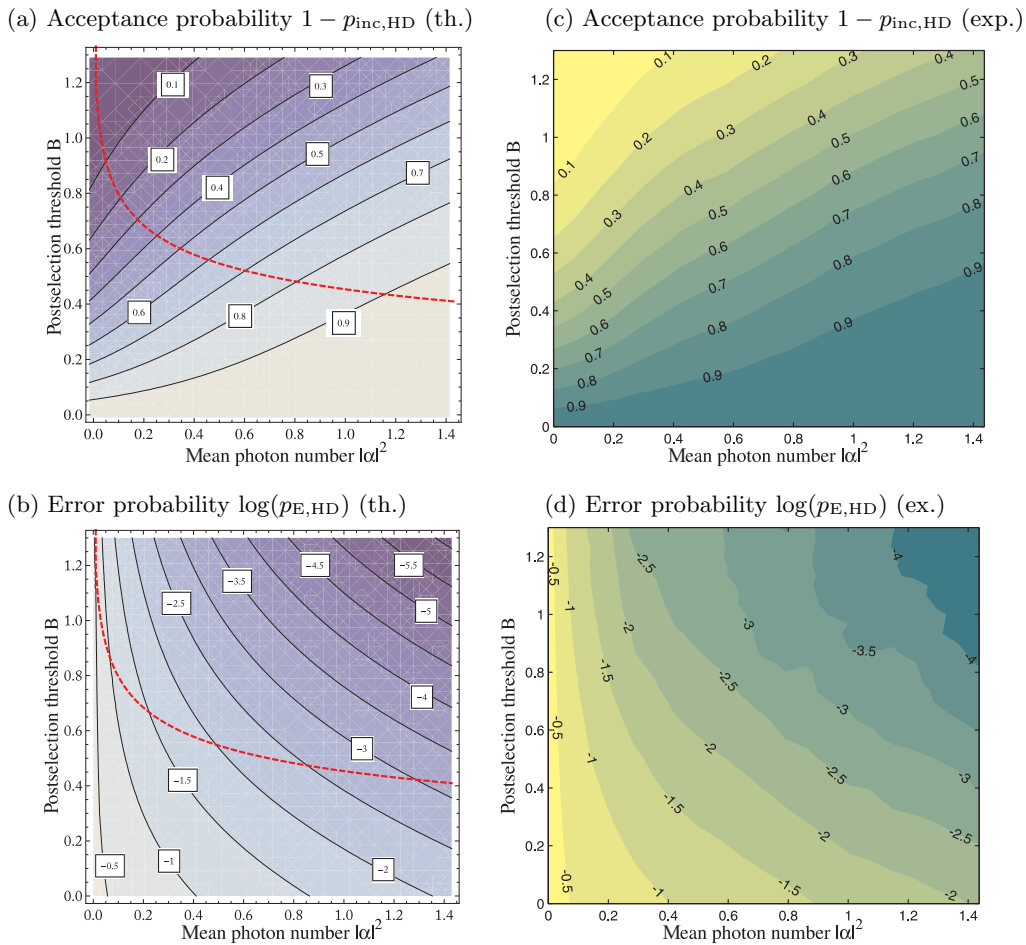}}
\\ [-12.6cm]
(a) Acceptance probability $1-p_{\mathrm{inc,HD}}$ (th.)&\hspace{0.8cm}(c) Acceptance probability $1-p_{\mathrm{inc,HD}}$ (exp.)\\ [6.1cm]
(b) Error probability $\mathrm{log}(p_{\mathrm{E,HD}})$ (th.) &\hspace{0.8cm}(d) Error probability $\mathrm{log}(p_{\mathrm{E,HD}})$ (ex.)\\ [5.7cm]
\end{tabular}
\caption{\label{HDrates} (a) The acceptance probability for ideal homodyne detection is depending on the signals mean photon number $|\alpha|^2$ and the postselection threshold $B$.  A dashed line shows for which postselection threshold $B$, the homodyne detection and USD have equal acceptance probability. (b) The error probabilities of ideal homodyne detection is shown in a logarithmic contour plot. The dashed line shows the error rate of homodyne detection for success rates equal to USD. (c) Experimentally measured acceptance probability. (d) Experimentally measured error probabilities. In (c) and (d), we corrected for the quantum efficiency of the receiver.}
\end{figure*}

In the following section, the receiver schemes are demonstrated with the experimental setup shown in Fig.~\ref{SetupRaw}. It consists of a preparation stage and two different receiver stages; the PNR receiver and a homodyne receiver. Our source is a grating stabilized CW diode laser at $810\:\mathrm{nm}$ (Toptica DL100). After passing a fiber mode cleaner (FMC), the linearly polarized beam is split asymmetrically in two parts to serve as a local oscillator of the homodyne receiver (LO) and an auxiliary oscillator (AO) for state preparation and displacement in the PNR receiver scheme. The signal state (S) is generated in a polarisation mode orthogonal to the auxiliary mode using an electro-optical modulator: The field amplitude of the auxiliary mode is coherently transfered into the signal polarisation and the excitation is controlled by the input voltage of the modulator. Note that the auxiliary oscillator remains in the polarisation mode orthogonal to the signal mode thus propagating along with the signal. After splitting the signal on a 50/50 beam splitter, two identical signal states (either $|\alpha\rangle^{\otimes 2}$ or $|-\alpha\rangle^{\otimes 2}$) are produced and subsequently directed to the two detection schemes. 

The signal states are generated in time windows of $\tau=800\:\mathrm{ns}$ with a repetition rate of $100\:\mathrm{kHz}$. This was done by applying a constant voltage across the modulator during the measurement time. The birefringence induced by the EOM's input voltage causes a variable coupling between the S and AO mode similar to a variable beam splitter. We can therefore tune the signal amplitude continuously. 

This modulation scheme is in contrast to the commonly used sideband modulation approach in experiments with homodyne detectors, where a RF modulation is applied to the modulator to create a pair of frequency sideband modes which defines the quantum state. Since the APD is not capable of selecting a specific pair of sideband modes, such a sideband approach cannot be used when the homodyne detector is used in conjunction with an APD. The quantum states are therefore defined as a pulse in the temporal frame of the local oscillator. The exact pulses measured by the two detectors are not identical as they have different frequency response. The effect of the detector response is described after the detailed description of the detection schemes.

We carefully characterise the prepared input signal and verify that the excess noise added to the quadrature by the signal preparation is only $5\cdot10^{-3}$ shot noise units (see Ref.~\cite{Wittmann2008c}). Such has purity is achieved by attenuation of the laser (the carrier) down to the single photon level thereby minimizing the thermal fluctuations at low frequencies prevailing the diode laser. 

At the homodyne receiver the signal interferes with the local oscillator, the two resulting outputs are detected and the difference current is produced. This yields an integrated quadrature value for each signal pulse. The detected signal of the homodyne detector is filtered with a 7-pole Chebyshev low pass filter from DC to 10MHz, and subsequently, the signal is sampled with 20MS/s. For a single pulse, the number of produced samples was therefore 16. These data was then averaged and thus resulting in a single quadrature measurement for a 800ns pulse. The technical noise at low frequencies was removed by correcting for the base line shift occurring between consecutive signal states. The phase of the signal relative to the LO is estimated by sending a number of bright calibration pulses along with the signal pulses. Subsequently, the measurements are accepted or discarded according to the estimated phase, i.e. they are only accepted if the measurement was performed along the signal encoding quadrature. This substitutes a technically demanding phase locking method. A drawback is the increasing measurement time. The overall quantum efficiency of the homodyne receiver amounts to $\eta_{\mathrm{hom}}=85.8\%$; the interference contrast to the local oscillator is $96.6\pm0.1\%$ and the PIN-diode quantum efficiency is $92\pm3\%$. The electronic noise level is more than $23\:\mathrm{dB}$ below the shot noise level.

The PNR receiver is composed of a displacement operation and a fiber coupled avalanche photo diode (APD) operating in an actively gated mode to circumvent a dark count event and thus a dead time at the time the pulse is arriving. During the measurement time, the APD works as a primitive photon number resolving detector, as used in~\cite{Banaszek1999}. The quantum states are subsequently categorized according to the respective result of the photon number measurement, thus implementing the POVMs  $\hat\Pi_1=|0\rangle\langle 0|$,   $\hat\Pi_?=\sum_{n=1}^{m}{|n\rangle\langle n|}$ and $\hat\Pi_2=\hat I -\hat\Pi_1-\hat\Pi_?$. In contrast to the displacement operation depicted in Fig.~\ref{schemes}(b) where two spatially separated modes interfere on a beam splitter, in our setup the two modes (the auxiliary and the signal modes) are in the same spatial mode but have different polarisation modes (Fig.~\ref{schemes}(c)). The interference (and thus the displacement) is therefore controlled by a modulator and a polarizing beam splitter. This method facilitates the displacement operation and yields a very high interference contrast of 99.6\%. A high extinction ratio is of high importance as the mis-matched part of the auxiliary might impinge onto the APD and cause false detection events.  Such false counts can be detrimental to the discrimination task for receivers with low error rate and especially if the signal amplitude is relatively large. 

The detection efficiency of the scheme is estimated to $\eta_{\mathrm{on/off}}=55\%$, including the transmission coefficient of the modulator, the polarisation optics and the fiber of $89.1\%$ as well as the quantum efficiency of the APD of $63\pm3\%$. The latter efficiency was estimated by the APD click statistic for an input coherent state that was calibrated by the homodyne receiver. An optical isolator is used between the two detection schemes to prevent back scattering of the LO to the APD.

\begin{figure}%
\includegraphics[width=0.85\columnwidth]{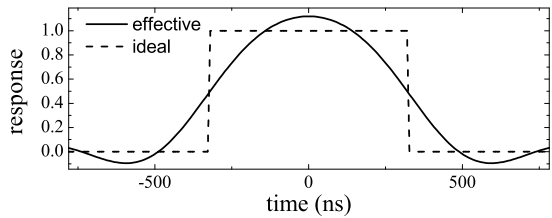}%
\vspace{-0.3cm}
\caption{\label{modeoverlap}Comparison between the effective and ideal response function of the homodyne receiver}%
\end{figure}

As the two detectors have different response, they do not measure the exact same temporal mode. However, in the following we show that the two different temporal modes are nearly identical possessing an overlap of about 95\%. In the experiment, we probe the optical mode $\hat a(t_m) =1/T\int_{t_m-T/2}^{t_m+T/2} \hat a(t) \mathrm{d}t$. The APD is a broadband detector, and by neglecting the electronic jitter noise and the dead time, the APD detects the following mean photon number during the measurement time T,
\begin{equation}
\hat n(t_m) = \frac{1}{T}\int_{t_m-T/2}^{t_m+T/2} \hat a^\dagger(t) \hat a(t) \mathrm{d}t
\label{eq:APDmode}
\end{equation}
On the other hand, the homodyne detector has a finite detector bandwidth which means that the quadrature measurement, $\hat X_{sa}$, at time $t$ depends on the detector response $G(\tau)$. A single sample is thus described by~\cite{Ou1995}
\begin{equation}
\hat X_\mathrm{sa}(t)  = \int \mathrm{d}\tau G(\tau) \hat X(t-\tau)
\label{eq:HDmode}
\end{equation}
where $G(\tau)$ is determined by the detectors frequency response $G(\omega)$ and accounts for the mean of $\hat X_{sa}$ over the measurement time T. The time-averaged measurement $\hat X_{av}$ at the time $t_m$ can then be written as
\begin{eqnarray}
\hat X_\mathrm{av}(t_m)&=& \frac{1}{T}\int_{t_m-T/2}^{t_m+T/2}\mathrm{d}t \int \mathrm{d}\tau G(\tau) \hat X(t-\tau)\\ \nonumber
&=& \frac{1}{T}\iint\mathrm{d}t \mathrm{d}\tau \mathrm{rect}(\frac{t-t_m}{T}) G(t-\tau) \hat X(\tau)\\ \nonumber
&=& \frac{1}{T}\int\mathrm{d}\tau\hat X(\tau)\int\mathrm{d}t  \mathrm{rect}(\frac{t-t_m}{T}) G(t-\tau) \\ \nonumber
&\mathrel{\widehat{=}}&\int \mathrm{d}\tau G_\mathrm{eff}(t_m-\tau) \hat X(\tau)
\label{eq:effG}
\end{eqnarray}
 where the measurement is described by the effective response function $G_\mathrm{eff}$. We compare this function to the ideal mode $\hat X_\mathrm{ideal}(t_m)= \frac{1}{T}\int_{t_m-T/2}^{t_m+T/2}\mathrm{d}t \hat X(t)$ as illustrated in Fig.~\ref{modeoverlap}. We estimate the mode overlap of the ideal and the effective mode with the normalized cross-correlation function $g_{12}=\langle \hat X_\mathrm{av}\hat X_\mathrm{ideal}\rangle/\sqrt{|\hat X_\mathrm{av}|^2|\hat X_\mathrm{ideal}|^2}$ (which is related to the visibility for the interference of two partially coherent waves~\cite{Saleh2007}). For our detector the cross-correlation is 95.5\,\%, and thus the similarity between the temporal mode seen by the APD and the one seen by the homodyne detector is about 95\%.

\begin{figure}
\begin{tabular}{l}
\centerline{\includegraphics[width=8cm]{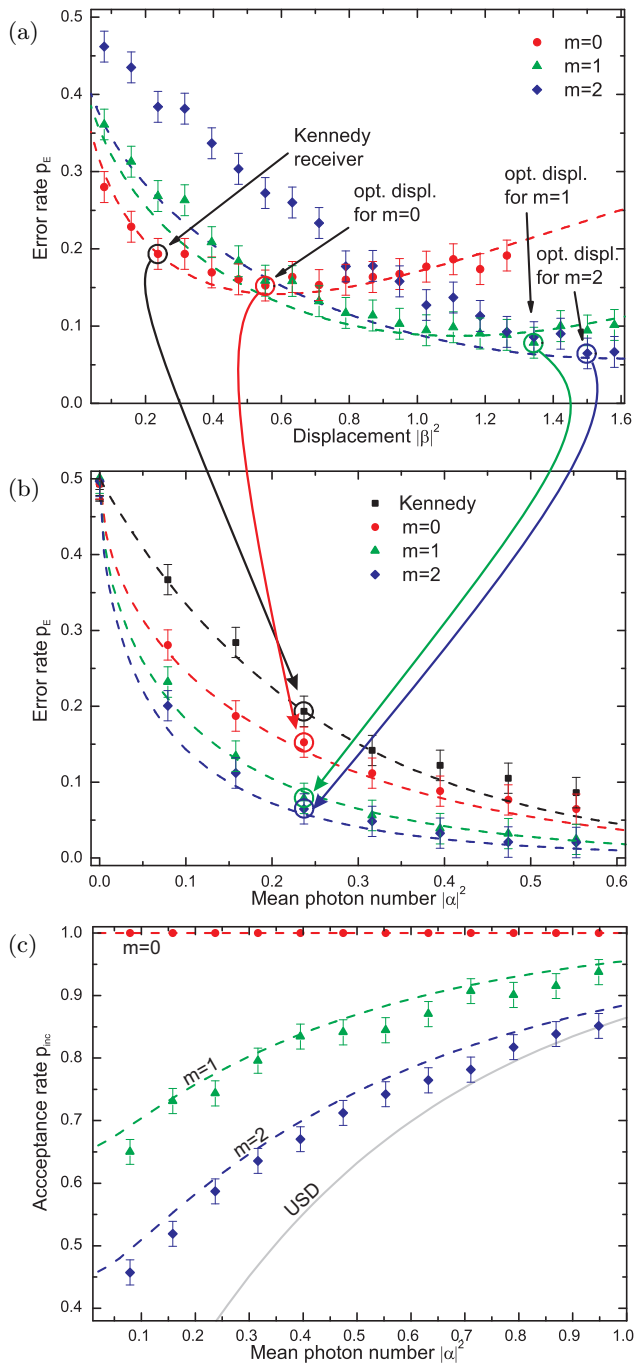}} \\ [-18cm]
(a)\\[5.7cm]
(b)\\ [5.7cm]
(c)\\ [5.4cm]

\end{tabular}
\caption{\label{Drates} (a) Experimental data on the effect of the displacement $\beta$ on the error rate $p_\mathrm{E,PNR}$ for a given signal amplitude $|\alpha|^2=0.24$ (corrected for quantum efficiency) and different number of dropped results $m$. Theoretically predicted error probabilities for the receiver without imperfections are shown with dashed lines. Error rates at the optimal displacement $\beta_\mathrm{opt}$ for different receivers are marked. (b) Error rates for varying $m$ and optimized $\beta_\mathrm{opt}$. Error bars reflect the standard deviations of repeated measurements, which are larger than the statistical errors. Experimental data is compared to ideal receivers (dashed lines). (c) Experimental data for acceptance rates(points) and theoretical predictions (dashed lines). Below the curve for an optimal USD device (grey), states can be discriminated without error in principle. We corrected for the quantum efficiencies of the receiver.}
 
\end{figure}

We proceed by describing the procedure of the discrimination task. A PC controls the preparation of the states and the displacement in the PNR receiver by modulating two electro optical modulators. Simultaneously it acquires the homodyne and APD detection outcomes during the pulse sequence. An example of such a sequence is shown in Fig.~\ref{SetupRaw}, where we show the voltages applied to the amplifiers, the quadrature values and the recorded number of counts per measurement time. The outcomes of the homodyne receiver within the interval $[-B,B]$ are considered as inconclusive results. If the value is outside the interval and positive we guess $|\alpha\rangle$ and if the value is outside the interval and negative we guess $|-\alpha\rangle$. For the outcomes of the PNR receiver, we use the hypothesis that if the outcome is larger than $m$, we guess $|\alpha \rangle$, if it is zero, we guess $|-\alpha\rangle$, otherwise it is an inconclusive result. The error probability is therefore found by adding up all the false detections and relate it to the number of pulses that were accepted. The acceptance probability is the ratio of pulses that were accepted to the total number of pulses.

The theoretical prediction for the acceptance probability $1-p_\mathrm{inc,HD}$ and the error probability $p_\mathrm{E,HD}$ are shown in Fig.~\ref{HDrates}(a) and (b) respectively. It is shown, that for increasing signal amplitudes $|\alpha|^2$ the error probability $p_\mathrm{E,HD}$ drops and that an increase of the postselection threshold $B$ leads to a decreasing error probability on the expense of an increase in the probability for inconclusive results. An advantage of the homodyne receiver is the smooth dependence between postselection threshold and error rate. This allows one to chose exactly the error rate desired for a specific application. 
For example in quantum key distribution the amount of mutual information between sender, adversary and receiver can be easily adjusted through the postselection threshold~\cite{Silberhorn2002}. The receivers performance is completely characterized by the error and acceptance rates. In the figures, we introduced a red dashed line, where the condition $p_{\mathrm{inc,HD}}=p_{\mathrm{inc,USD}}$ is met, with the probability of inconclusive results in a perfect USD measurement $p_{\mathrm{inc,USD}}$. This means, an error-free but probabilistic discrimination is in principle possible above this curve.

The experimental results for the acceptance and the error probability of the homodyne receiver are shown in Fig~\ref{HDrates}(b) and (d) respectively. The contour plots are generated from signals with 21 different amplitudes (with linearly increasing the mean photon number) and calculated for 41 postselection thresholds. We find very good agreement of theory and experiment with only minor deviations for very low error probabilities.

The PNR receiver is demonstrated for $m=0$, $1$ and $2$. In Fig.~\ref{Drates}(a), the dependence of the error probability on the displacement $\beta$ for a fixed signal amplitude is illustrated.  We find that for any $m$ the displacement can be optimized such that the experimentally measured error rates reach a minimum. The optimal displacement is higher for higher $m$ and the minimum error rate after this optimization of the displacement is lowered for increasing $m$. When compared to the theoretical predictions, the experimental data fits well in the region of the minima, while the experimental imperfections dominate in the region of smaller displacement. We could also observe this for $m>2$. 

We marked four data points in the plot. From left to right, they represent the error rates associated with the Kennedy receiver (black) (which is an early receiver for the minimum error discrimination~\cite{Kennedy1973} without optimized displacement, i.e. $\beta=\alpha$), the optimized displacement receiver with $m=0$ (red), and the PNR receivers with $m=1$ and $2$ (green and blue). The error rates for varying amplitudes are plotted in Fig.~\ref{Drates}(b). We find a maximal reduction of the error rate by a factor of $3.5$ going from $m=0$ (deterministic scheme) to $m=2$ (probabilistic scheme) at the signal amplitude $|\alpha|^2=0.47$. The corresponding penalty on the acceptance rates and the comparison with the theoretical predictions for the acceptance probability are shown in Fig.~\ref{Drates}(c).

\begin{figure}
\begin{tabular}{l}
\centerline{\includegraphics[width=8cm]{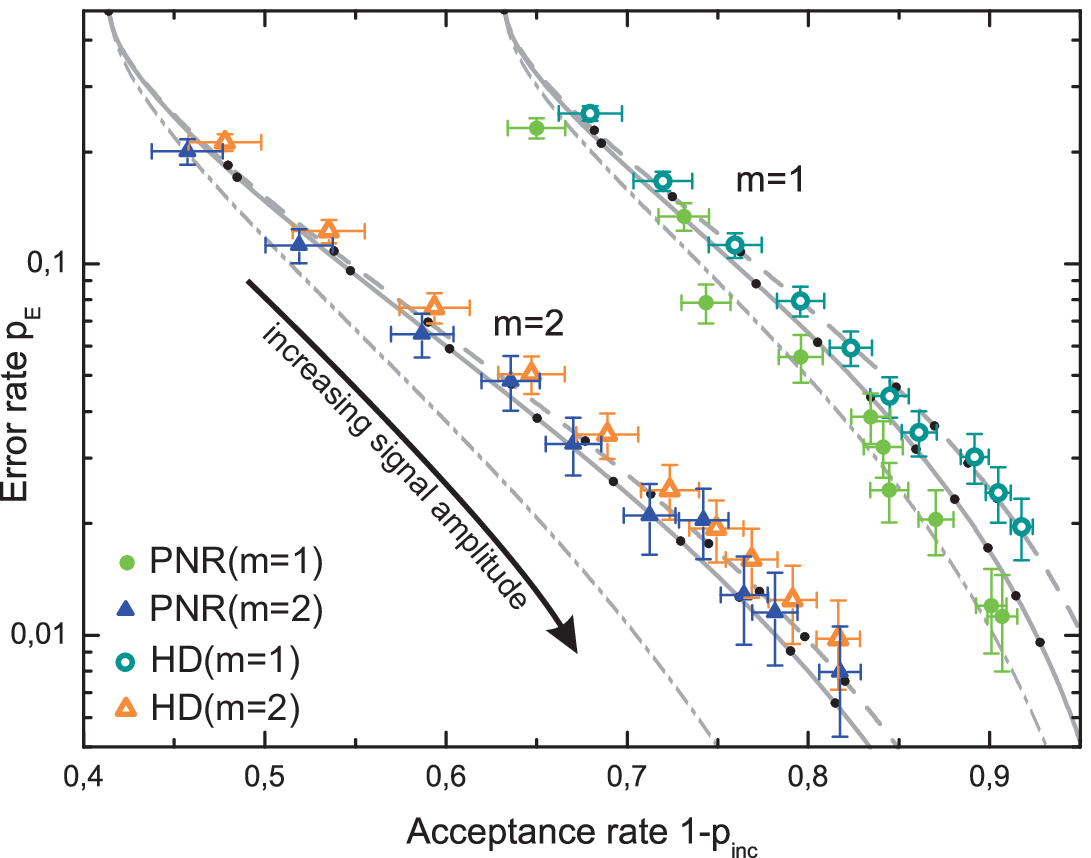}}\\
\centerline{\includegraphics[width=8cm]{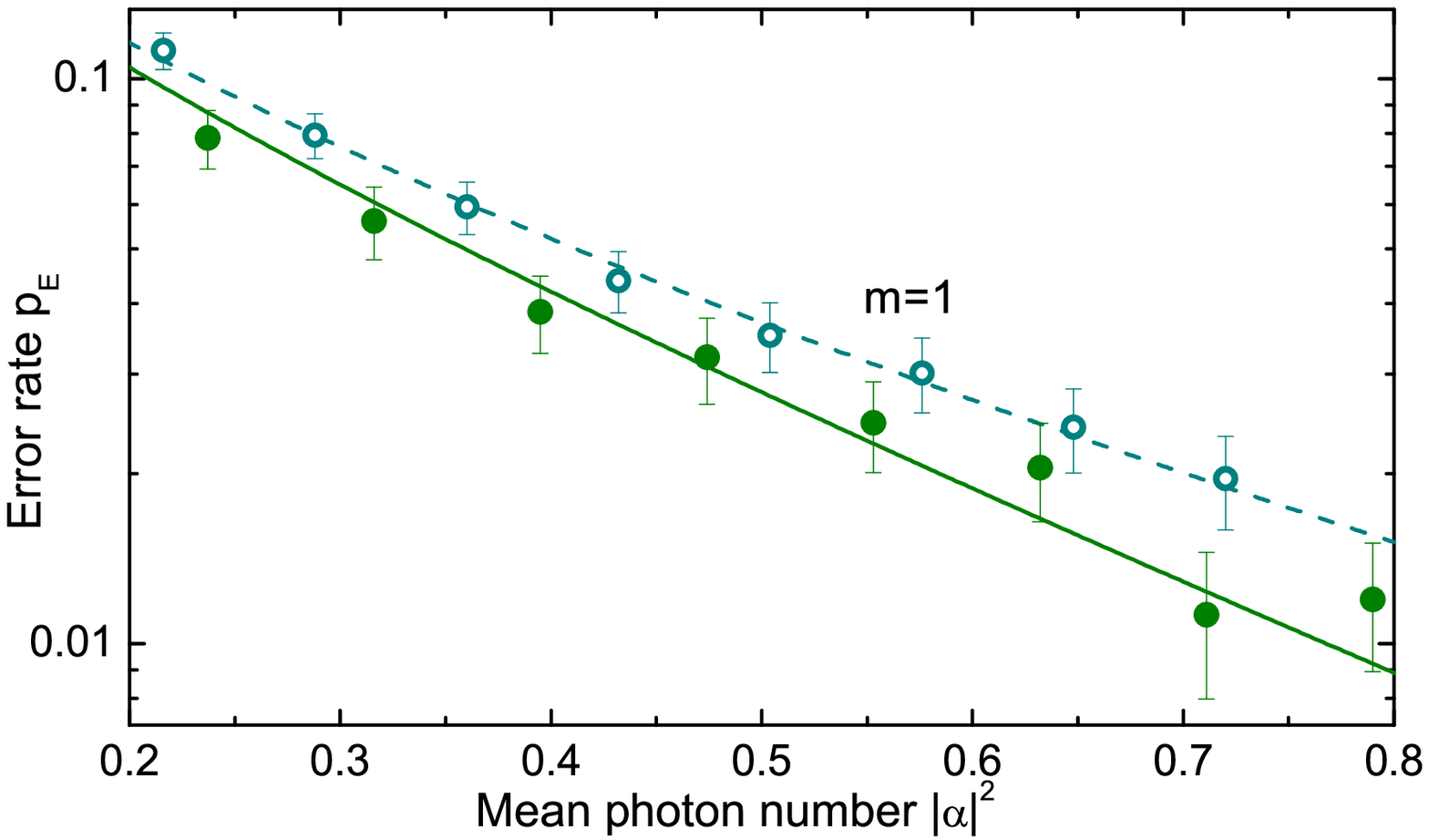}}\\[-11.3cm]
(a)\\[5.9cm]
(b)\\ [4.5cm]
\end{tabular}
\caption{\label{comprates} (a) Experimental error rates and acceptance rates for the two receiver schemes are compared. For this comparison the success rate of both schemes is fixed to the one, that is theoretically reached by the PNR receiver. Experimental data is shown for PNR receivers with $m=1$ and $2$ (filled circles and triangles) and homodyne receiver (open circles and triangles). Additionally, theoretical predictions for the homodyne receiver (grey dashed line), the PNR receiver (solid line) and the optimal intermediate measurement (dotdashed lines) are shown. The mean photon number is varied for the different receivers. The PNR receiver again outperforms the homodyne receiver and we find a relatively good agreement of experimental data and theoretical predictions. (b) Error rate for various signal amplitudes. The PNR receiver surpasses the homodyne receiver. Statistical error bars show standard deviation of the random process. We corrected for the quantum efficiencies of the receivers.}
 
\end{figure}

Both detection schemes are compared to each other in Fig.~\ref{comprates}. We find that both receivers show the expected behavior. Especially for $m=1$, it is obvious that the PNR receiver outperforms the homodyne receiver for several data points. 

In the following, we discuss the limitations of the different schemes. The quantum efficiency of homodyne detection is partly limited by the PIN-diode efficiency and partly by the mode-matching efficiency at the homodyne's beam splitter. For special made PIN diodes, the efficiency can reach nearly 100\%, and the beam splitter mode matching efficiencies beyond 99\% have been reported. The efficiency of the PNR is mainly limited by the quality of the displacement operation and the efficiency of the avalanche photo diode (APD). For higher values of m, we also find that also the probability of false detection events becomes important (see $m {=} 2$ in Fig.~\ref{comprates}(a)). We used a commercial available APD, but the development of photon number resolving detectors with very high quantum efficiency is rapidly progressing(see~\cite{Worsley2009} for a detailed list).

\section{Conclusion}
\label{comp}

In this paper we have thoroughly investigated two different probabilistic receivers of binary-encoded optical coherent states; the homodyne detector and the displacement controlled photon number resolving detector. These receivers yield inconclusive as well as error affected results and we have carefully conducted a detailed study of the relation between these two outcomes. Furthermore, we found, theoretically, that the homodyne detector is the optimal Gaussian receiver for minimizing the errors for a fixed probability of inconclusive results. Experimentally, we have implemented both receivers and through comparison we found that the performance of the new displacement controlled PNR is better than the homodyne receiver. 

The new receiver is thus a promising alternative to the commonly used homodyne receiver. We find several advantages of the new scheme. For example, the power of the displacement beam is normally much lower than the power of the local oscillator beam required for homodyne detection, and thus less power is injected into the communication channel (e.g. an optical fibre). This has the obvious benefit of lowering the power consumption in the fibre but it also lowers the risk of scattering of auiliary photons into the signal state as such scattering mechanism is proportional to the power. We also note that instead of performing the displacement operation at receiving station, it can be already implemented at the sending station. This would completely remove the necessity of a phase reference. Finally, we note that the quadrature measurement can be also performed using a displacement operation followed by a single intensity detector similar to the setup of the PNR detector. However, in the former case the displacement must be macroscopic such that the quadratures are measured instead of the photon properties.  

The new receiver is fundamentally different from the more commonly used and technically mature homodyne detector. Whether the PNR detector will be the future choice in real life implementations of binary detectors will depend on the future technical progress of the PNR detector technology. A future option is also to use both detector schemes in a receiving station, where the proper detection scheme is chosen according to the currently needed property, such as speed, low noise and the capability of performing a full tomography of the state~\cite{Qi2007d}.

The work has been supported by the EU project COMPAS and Lundbeckfonden(R13-A1274).



\end{document}